\documentclass[aps,prd,preprint,superscriptaddress,showpacs,floatfix,nobibnotes,nofootinbib]{revtex4}
\usepackage{mathrsfs}

\usepackage{bm}
\usepackage{graphicx}
\usepackage{longtable}
\usepackage{amsmath}
\usepackage{amssymb}

\begin{document}

\title{Development of Bethe-Salpeter theory for dealing with unstable system}

\author{Xiaozhao Chen}\email{chen_xzhao@sina.com}
\email[corresponding author]{} \affiliation{Department of Fundamental Courses, Shandong University of Science and Technology, Taian, 271019, China}

\author{Xiaofu L\"{u}}
\affiliation{Department of Physics, Sichuan University, Chengdu, 610064, China}
\affiliation{Institute of Theoretical Physics, The Chinese Academy of Sciences, Beijing 100080, China}
\affiliation{CCAST (World Laboratory), P.O. Box 8730, Beijing  100080, China}

\date{\today}

\begin{abstract}
In the framework of relativistic quantum field theory, the solution of homogeneous Bethe-Salpeter equation for two-body bound state can not describe unstable system, so we develop Bethe-Salpeter theory to investigate resonance which is regarded as an unstable two-body system. Based on Bethe-Salpeter wave function, we consider the time evolution of two-body bound state determined by the total Hamiltonian. The total matrix element for arbitrary decay channel is expressed in terms of the Heisenberg picture, and Mandelstam's approach is generalized to calculate the matrix element between bound states with respect to arbitrary value of the final state energy. Some innovations to Feynman diagram are made so that the key features of dispersion relation can be more clearly exhibited. This new resonance theory in quantum field theory is applied to investigate exotic particle which is considered as an unstable meson-meson molecular state.
\end{abstract}

\pacs{12.40.Yx, 14.40.Rt, 12.39.Ki}

%\keywords{Resonance; Bethe-Salpeter equation; Bound state matrix element }

%\bigskip
\maketitle

\newpage

\parindent=20pt

\section{Introduction}
\label{sec:intro}

Many exotic particles have been discovered in experiment and many possible alternative interpretations beyond quark-antiquark state have been proposed in theory \cite{ms:Swanson,ms:Torn,liu,ds:Maian1,ds:Maian2,ts:Ebert}. Among these interpretations, homogeneous Bethe-Salpeter (BS) equation is frequently used to investigate the internal structure of exotic particles which are considered as two-body bound states \cite{ms:Branz,mypaper4,mypaper6,mypaper7,Msigma2}. In quantum field theory, homogeneous BS equation is a nonperturbative method \cite{BSE:Roberts3,BSE:Roberts4,BSE:Roberts5}, which should be only applied to deal with two-body bound state in the strict sense because the solution of homogeneous BS equation can not contain the contribution from decay channels. However, in experiments exotic particles are resonances, so these exotic particles are unstable states which should not be completely treated as stationary two-body bound states and it is more reasonable to regard exotic resonances as unstable two-body systems. More importantly, present field theory seldom takes into account the difference between the internal structure of stable two-body bound state and the one of unstable two-body system, and the theory describing internal structure of resonance has not been constructed in field theory. In this work, two-body bound state is a stable composite particle and resonance is regarded as an unstable composite particle. We develop homogeneous BS theory to describe the internal structure of resonance in the framework of relativistic quantum field theory and illustrate this new theory based on BS equation for exotic meson resonance.

In our previous works \cite{mypaper4,mypaper6,mypaper7}, exotic particles were considered as meson-meson bound states. Solving homogeneous BS equations for meson-meson bound states, we obtained masses and BS wave functions. The mass of meson-meson bound state was regarded as mass of exotic meson resonance and the correction for energy level of molecular state due to decay channels has not been considered \cite{mypaper4,mypaper6,mypaper7}. As well-known, all decay channels of resonance should contribute to its physical mass. However, the technique about the dynamics of coupled channels based on Schr\"{o}dinger wave function still remains in nonrelativistic case \cite{couchan1,couchan2,couchan3}. In relativistic quantum field theory, the Schr\"{o}dinger wave function is not a useful quantity to work with, in view of its non-Lorentz-invariant character. So far, the technique about the dynamics of coupled channels based on covariant Bethe-Salpeter wave function has not been established. Therefore, it is necessary and important to seek a development of homogeneous BS theory for dealing with the dynamics of coupled channels in the framework of relativistic quantum field theory.

In this paper, exotic meson resonance is considered as an unstable meson-meson molecular state. Based on BS wave function for meson-meson bound state, we can provide a description for the prepared state and then study the time evolution of meson-meson molecular state determined by the total Hamiltonian. Using dispersion relation, the Heisenberg picture and Mandelstam's approach, we obtain the correction for energy level of resonance and then the physical mass is used to calculate its decay width. An innovative Feynman diagram is introduced, in which the key features of dispersion relation can be exhibited clearly.

\section{Time evolution of BS wave function used to study energy level and decay width of resonance}
Let us begin with the interaction Lagrangian for the coupling of light quark fields to light meson fields as in effective theory at low energy QCD. According to the effective theory at low energy QCD, non-vanishing vacuum condensate causes the spontaneous breaking of chiral symmetry, which leads to the appearance of Goldstone bosons \cite{QTFII}. At low energy QCD, the effective interaction Lagrangian can be regarded as Lagrangian for the interaction of light mesons with quarks \cite{mypaper6}
\begin{equation}
\begin{split}\label{Lag}
\mathscr{L}_I=ig_0\bar{\mathcal{Q}}\gamma_5\mathbb{P}\mathcal{Q}+ig'_0\bar{\mathcal{Q}}\gamma_\mu \mathbb{V}_\mu\mathcal{Q}+g_\sigma\bar{\mathcal{Q}'}\mathcal{Q}'\sigma,
\end{split}
\end{equation}
where $\bar{\mathcal{Q}}=(\bar u,\bar d,\bar s)$, $\bar{\mathcal{Q}'}=(\bar u,\bar d)$, $g$ represents the corresponding meson-quark coupling constant, $\mathbb{P}$ and $\mathbb{V}$ are the octet pseudoscalar and nonet vector meson matrices, respectively. From this Lagrangian, we have investigated the light meson interaction with quarks in heavy mesons and obtained the interaction of heavy meson with light meson through the heavy meson form factor \cite{mypaper4,mypaper5}.

Using path integrals, one can obtain a homogeneous integral equation for arbitrary bound state composed of two mesons. In symbolic notation homogeneous BS equation may be written as
\begin{equation}\label{BSE}
\begin{split}
(S^{(1)-1}S^{(2)-1}+\mathcal{V})\chi=0,
\end{split}
\end{equation}
where $\chi$ represents BS wave function, the kernel $\mathcal{V}$ is the sum of all irreducible graphs, $S^{(1)}$ and $S^{(2)}$ represent meson propagators, respectively. Solving this BS equation, one can obtain the mass $M_0$ and BS wave function $\chi_P(x_1,x_2)$ for this meson-meson bound state with momentum $P=(\mathbf{P},i\sqrt{\mathbf{P}^2+M_0^2})$. We emphasize that the kernel $\mathcal{V}$ in homogeneous BS equation expressed as Eq. (\ref{BSE}) is defined in two-body channel so $\mathcal{V}$ is not complete interaction. The kernel in homogeneous BS equation (\ref{BSE}) plays a central role for making two-body system to be a stable bound state, and the solution of homogeneous BS equation (\ref{BSE}) should only describe bound state.

Since resonance decays spontaneously into other particles, we can suppose that at the times $t_1=0$ and $t_2=0$ this unstable state has been prepared to decay. This prepared state (ps) can be described by the ground-state BS wave function which has the form
\begin{equation}
\begin{split}\label{BSWFT0}
\mathscr{X}^{\text{ps}}_a&=\chi_P(\mathbf{x}_1,t_1=0,\mathbf{x}_2,t_2=0)=\frac{1}{(2\pi)^{3/2}}\frac{1}{\sqrt{2E(P)}}e^{i\mathbf{P}\cdot(\eta_1\mathbf{x}_1+\eta_2\mathbf{x}_2)}\chi_P(\mathbf{x}_1-\mathbf{x}_2),
\end{split}
\end{equation}
where $E(p)=\sqrt{\mathbf{p}^2+m^2}$ and $\eta_1+\eta_2=1$. Then the time evolution of this system determined by the total Hamiltonian $H$ has the explicit form
\begin{equation}
\begin{split}\label{timeevo}
\mathscr{X}(t)=e^{-iHt}\mathscr{X}^{\text{ps}}_{a}=\frac{1}{2\pi i}\int_{C_2}d\epsilon e^{-i\epsilon t}\frac{1}{\epsilon-H}\mathscr{X}^{\text{ps}}_{a},
\end{split}
\end{equation}
where $G(\epsilon)=(\epsilon-H)^{-1}$ is the Green's function and the contour $C_2$ runs from $ic_r+\infty$ to $ic_r-\infty$ in energy-plane. The positive constant $c_r$ is sufficiently large that no singularity of $(\epsilon-H)^{-1}$ lies above $C_2$. The Green's function can be represented by scattering matrix \cite{GreenFun}
\begin{equation}
\begin{split}\label{Gaepsilon}
G_{aa}(\epsilon)=(\chi^{\text{ps}}_{a},G(\epsilon)\chi^{\text{ps}}_{a})=\frac{1}{\epsilon-M_0-(2\pi)^3T_{aa}(\epsilon)},
\end{split}
\end{equation}
where $\chi^{\text{ps}}_{a}$ represents $(2\pi)^{-3/2}[2E(P)]^{-1/2}\chi_P(\mathbf{x}_1-\mathbf{x}_2)$ in Eq. (\ref{BSWFT0}). The proof of Eq. (\ref{Gaepsilon}) has been given by Ref. \cite{GreenFun}. This work will give $T_{aa}(\epsilon)$ in the framework of relativistic quantum field theory. In field theory the operator $T(\epsilon)$ is just the scattering matrix with energy $\epsilon$, and $T_{aa}(\epsilon)$ is the $T$-matrix element between two bound states, which should be defined as $\langle a~\text{out}|a~\text{in}\rangle=\langle a~\text{in}|a~\text{in}\rangle-i(2\pi)^4\delta^{(4)}(P-P)T_{aa}(\epsilon)$.

Because of the analyticity of $T_{aa}(\epsilon)$, we define
\begin{equation}
\begin{split}\label{Taepsilon}
T_{aa}(\epsilon)=\mathbb{D}(\epsilon)-i\mathbb{I}(\epsilon),
\end{split}
\end{equation}
where $\epsilon$ approaches the real axis from above, $\mathbb{D}$ and $\mathbb{I}$ are the real and imaginary parts, respectively. When there is only one decay channel, we can use the unitarity of $T_{aa}(\epsilon)$ to obtain \cite{GreenFun}
\begin{equation}
\begin{split}\label{Taepsilon1}
2\mathbb{I}(\epsilon)=\sum_b(2\pi)^4\delta^{(3)}(\mathbf{P}_b-\mathbf{P})\delta(E_b-\epsilon)|T_{ba}(\epsilon)|^2,
\end{split}
\end{equation}
where $P_b=(\mathbf{P}_b,iE_b)$ is the total energy-momentum vector of all particles in the final state and the $T$-matrix element $T_{ba}(\epsilon)$ is defined as $\langle b~\text{out}|a~\text{in}\rangle=-i(2\pi)^4\delta^{(3)}(\mathbf{P}_b-\mathbf{P})\delta(E_b-\epsilon)T_{ba}(\epsilon)$. The delta-function in Eq. (\ref{Taepsilon1}) means that the energy $\epsilon$ in scattering matrix is equal to the total energy $E_b$ of the final state, and $\sum_b$ represents summing over momenta and spins of all particles in the final state. For $E_b=\epsilon$, we also denote the total energy of the final state by $\epsilon$ and $\mathbb{I}(\epsilon)$ becomes a function of the final state energy. Using dispersion relation for the function $T_{aa}(\epsilon)$, we obtain
\begin{equation}
\begin{split}\label{disrel}
\mathbb{D}(\epsilon)=-\frac{\mathcal{P}}{\pi}\int_{\epsilon_M}^\infty \frac{\mathbb{I}(\epsilon')}{\epsilon'-\epsilon}d\epsilon'.
\end{split}
\end{equation}
The symbol $\mathcal{P}$ means that this integral is a principal value integral and the variable of integration is the total energy $\epsilon'$ of the final state. In order to obtain the real part $\mathbb{D}(\epsilon)$, we should calculate the function $\mathbb{I}(\epsilon')$ of value of the final state energy $\epsilon'$, which is an arbitrary real number over the real interval $\epsilon_M<\epsilon'<\infty$. As usual the momentum of initial bound state $a$ is set as $P=(0,0,0,iM_0)$ in the rest frame and $\epsilon_M$ denotes the sum of all particle masses in the final state.

Let us suppose that there are several decay channels and the final state $b$ may contain $n$ composite particles and $n'$ elementary particles in decay channel $c'$. From the Heisenberg field operator corresponding to elementary particle in Lagrangian, one can obtain the in-field and out-field by means of the Yang-Feldman equations. Asymptotically an elementary particle is represented by the in-field and out-field satisfying the corresponding free field equations of motion. Then the in-state and out-state consisting of elementary particle can be generated and the ordinary $S$-matrix element has been well defined. For the composite particle composed by elementary particles, it is clearly impossible to introduce a Heisenberg field operator corresponding to this composite particle. However, asymptotically composite particle can also be represented by the in-field and out-field satisfying the corresponding free field equations of motion \cite{ParticlesFields}, and then the in-state and out-state consisting of composite particle can also be generated. As far as the asymptotic condition is concerned, there is no clear distinction between elementary and composite particles in the final state $b$.

This work investigates the time evolution of unstable composite particle and the initial bound state $a$ must be different from the final composite particle, so we have to consider the internal structure of composite particle in the final state $b$. From Eq. (\ref{Taepsilon1}), we have
\begin{equation}
\begin{split}\label{Iepsilon'}
\mathbb{I}(\epsilon')=&\frac{1}{2}\sum_{c'}\int d^3Q'_1...d^3Q'_{n'}d^3Q_1...d^3Q_n(2\pi)^4\delta^{(4)}(Q'_1+...+Q_n-P^{\epsilon'})\sum_{\text{spins}}|T_{(c';b)a}(\epsilon')|^2,
\end{split}
\end{equation}
where $Q'_1...Q'_{n'}$ and $Q_1...Q_n$ are the momenta of final elementary and composite particles, respectively; $P^{\epsilon'}=(0,0,0,i\epsilon')$, $T_{(c';b)a}(\epsilon')$ is the $T$-matrix element with respect to $\epsilon'$, $\sum_{\text{spins}}$ represents summing over spins of all particles in the final state, and $\sum_{c'}$ represents summing over all open and closed channels. In Eq. (\ref{Iepsilon'}) the energy in scattering matrix is equal to the total energy $\epsilon'$ of the final state $b$, which is an arbitrary real number over the real interval $\epsilon_M<\epsilon'<\infty$. The mass $M_0$ and BS amplitude of initial bound state $a$ have been specified and the value of the initial state energy in the rest frame is a specified value $M_0$. From Eq. (\ref{Iepsilon'}), we have $\mathbb{I}(\epsilon')>0$ for $\epsilon'>\epsilon_M$ and $\mathbb{I}(\epsilon')=0$ for $\epsilon'\leqslant\epsilon_M$, which is the reason that the integration in dispersion relation (\ref{disrel}) ranges from $\epsilon_M$ to $+\infty$. Therefore, in order to obtain the real part $\mathbb{D}(\epsilon)$ due to all open and closed channels, we have to calculate $T_{(c';b)a}(\epsilon')$ with respect to value of the final state energy $\epsilon'$, which is an arbitrary real number over the real interval $\epsilon_M<\epsilon'<\infty$. We emphatically introduce the $T$-matrix element $T_{(c';b)a}(\epsilon')$ as follows.

In our theory $T_{(c';b)a}(\epsilon')$ must involve bound state, so this matrix element can not be calculated as an ordinary $S$-matrix element. Because it is impossible to introduce a Heisenberg field operator corresponding to composite particle, we define the in-field and out-field to represent composite particle. For example, we suppose that the initial bound state is a scalar composite particle composed of two elementary particles. Asymptotically this composite particle can be represented by in-field $\phi^{M_0}_{\text{in}}(X)$ satisfying the corresponding free field equation of motion \cite{ParticlesFields}
\begin{equation}
\begin{split}
(\square-M_0^2)\phi^{M_0}_{\text{in}}(X)=0,
\end{split}
\end{equation}
where $M_0$ is the mass of bound state, $X$ is the centre of mass coordinate and $X=\eta_1x_1+\eta_2x_2$. Then we can define the creation operator $a^{M_0\dag}_{\textbf{P},\text{in}}$ in $\phi^{M_0}_{\text{in}}$, which generates the in-state $|P~\text{in}\rangle=a^{M_0\dag}_{\textbf{P},\text{in}}|0\rangle$. Similarly, $n$ composite particles in the final state can be generated by $n$ out-fields, respectively, and the final state including $n$ composite particles is out-state $\langle Q_1...Q_n~\text{out}|$. Therefore, the initial bound state $a$ is $|P~\text{in}\rangle$ and the final state $b$ including $n$ composite particles and $n'$ elementary particles is $\langle Q'_1...Q'_{n'},Q_1...Q_n~\text{out}|$.

\begin{figure}[tbp] \centering
\includegraphics[trim = 0mm 10mm 0mm 0mm,scale=1,width=7cm,origin=c]{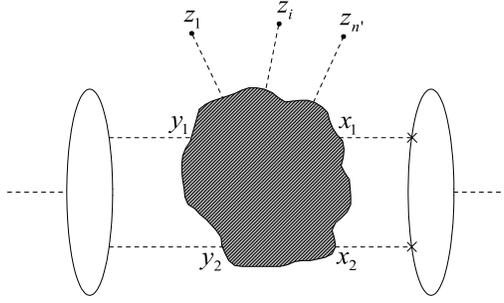}%trim=l b r t This option will crop the imported image by l from the left, b from the bottom, r from the right, and t from the top. Where l, b, r and t are lengths.
\caption{\label{Fig1} General matrix element between two bound states $\langle Q~\text{out}|T\phi(z_1)...\phi(z_i)...\phi(z_{n'})|P~\text{in}\rangle$ with respect to $\epsilon'$. The filled blob represents the two-particle irreducible Green's function, and the unfilled ellipses represent BS amplitudes. The crosses mean that the final state energy extends from $\epsilon_{M}$ to $+\infty$ while the initial state energy is specified, which is different from the traditional Feynman diagram.}
\end{figure}

Using the Heisenberg picture, we can obtain the total matrix element between a final state $\langle Q'_1...Q'_{n'},Q_1...Q_n~\text{out}|$ and a specified initial bound state $|P~\text{in}\rangle$
\begin{equation}
\begin{split}\label{Rmatrix}
-iR_{(c';b)a}(\epsilon')=&\langle Q'_1...Q'_{n'},Q_1...Q_n~\text{out}|P~\text{in}\rangle\\
=&i^{2n'}\int d^4z_1...d^4z_{n'}f_{Q'_1}^*(z_1)...f_{Q'_{n'}}^*(z_{n'})S_{z_1}'^{-1}...S_{z_{n'}}'^{-1}\\
&\times \langle Q_1...Q_n~\text{out}|T\phi(z_1)...\phi(z_i)...\phi(z_{n'})|P~\text{in}\rangle.
\end{split}
\end{equation}
Here these Heisenberg fields $\phi$ represent the boson fields appearing in Lagrangian. The functions $f$ are solutions to the corresponding free field equations of motion, and these $S'$ represent free boson propagators. In Eq. (\ref{Rmatrix}), we have used reduction formulae to deal with elementary particles and of great interest is the matrix element of a time-order product of Heisenberg field operators between bound states. Mandelstam's approach is a technique based on BS wave function for evaluating the general matrix element between bound states. Applying Mandelstam's approach, one can express the general matrix element between bound states in terms of BS wave functions and a two-particle irreducible Green's function, and the proof has been given by Ref. \cite{ParticlesFields}. In this work, Mandelstam's approach is generalized to evaluate the bound state matrix element with respect to arbitrary value of the final state energy $\epsilon'$
\begin{equation}
\begin{split}\label{Gmatrele}
\langle Q_1...Q_n&~\text{out}|T\phi(z_1)...\phi(z_i)...\phi(z_{n'})|P~\text{in}\rangle\\
=&\int d^4y_1d^4y_2...d^4y_{2n-1}d^4y_{2n}d^4x_1d^4x_2\\
&\times\bar\chi_{Q_1}(y_1,y_2)...\bar\chi_{Q_n}(y_{2n-1},y_{2n})\mathbb{T}(y_1...y_{2n};z_1...z_i...z_{n'};x_1,x_2)\chi_P(x_1,x_2),
\end{split}
\end{equation}
where $\mathbb{T}(y_1...y_{2n};z_1...z_i...z_{n'};x_1,x_2)$ is the two-particle irreducible Green's function, $\bar\chi$ and $\chi$ are BS wave functions for the final and initial bound states, respectively. The function $\mathbb{T}$ can, in principle, be evaluated by means of perturbation theory. It is necessary to emphasize that the general matrix element (\ref{Gmatrele}) is calculated with respect to $\epsilon'$, and the energy in $\mathbb{T}$ is equal to the final state energy $\epsilon'$, which is an arbitrary real number over the real interval $\epsilon_M<\epsilon'<\infty$. The initial state energy in the rest frame $M_0$ and BS amplitude of initial bound state have been specified. But the traditional Feynman diagram represents only the case that the value of the final state energy is just the value of the initial state energy $\epsilon'=M_0$ and can not represent closed channel. In this paper we introduce a Feynman diagram to represent arbitrary value of the final state energy, called \emph{extended Feynman diagram}, shown in Figure \ref{Fig1}. The crosses in extended Feynman diagram mean that the energy in $\mathbb{T}$ is equal to the final state energy $\epsilon'$ extending from $\epsilon_{M}$ to $+\infty$ while the energy $M_0$ of initial bound state is specified. When $\epsilon'=M_0$, the crosses in extended Feynman diagram disappear and the extended Feynman diagram becomes the traditional Feynman diagram. Removing delta-function factor $(2\pi)^4\delta^{(4)}(Q'_1+...+Q_n-P^{\epsilon'})$ in $R_{(c';b)a}(\epsilon')$, we obtain $T_{(c';b)a}(\epsilon')$.

To illustrate this, we imagine that the initial bound state ($MS$) is composed of two heavy vector mesons ($VM$ and $\overline{VM}$) and the final state contains a heavy meson ($HM$) and a light meson ($LM$). If a bound state with spin $j$ is created by two massive vector fields, its BS wave function can be defined as $\chi_{P(\lambda\tau)}^j(x_1,x_2)=\langle0|TA_\lambda(x_1)A^\dagger_\tau(x_2)|P,j\rangle$ and we have given the general form for this BS wave function $\chi_{\lambda\tau}^{j}(P,p)$ in the momentum representation, where $p$ is the relative momentum of two vector fields \cite{mypaper4,mypaper7}. This BS wave function should satisfy the equation
\begin{equation}\label{BSE1}
\begin{split}
\chi^{j}_{\lambda\tau}(P,p)=&-\int \frac{d^4p'}{(2\pi)^4}\Delta_{F\lambda\theta}(p_1')\mathcal{V}_{\theta\theta',\kappa'\kappa}(p,p';P)\chi^{j}_{\theta'\kappa'}(P,p')\Delta_{F\kappa\tau}(p_2'),
\end{split}
\end{equation}
where $\mathcal{V}_{\theta\theta',\kappa'\kappa}$ is the interaction kernel, $p_1'$ and $p_2'$ are the momenta carried by two vector fields, $\Delta_{F\lambda\theta}(p_1')$ and $\Delta_{F\kappa\tau}(p_2')$ are the propagators for the spin 1 fields. Owing to the effective interaction Lagrangian at low energy QCD (\ref{Lag}), we have to consider that the heavy meson is a bound state composed of a quark and an antiquark and investigate the interaction of light meson with quarks in heavy meson. Through the heavy meson form factor describing the heavy meson structure, we have obtained the interaction kernel between two heavy vector mesons ($VM$ and $\overline{VM}$) derived from one light meson ($\sigma$, $\omega$, $\rho$, $\phi$) exchange in Refs. \cite{mypaper4,mypaper3}. In our previous works \cite{mypaper4,mypaper5}, BS equation (\ref{BSE1}) has been solved and the mass $M_0$ and BS wave function $\chi_{\lambda\tau}^{j}(P,p)$ for the bound state composed of two heavy vector mesons have been obtained. In this paper, we are interested only in mass correction for molecular state and do not repeat the procedure for solving BS equation.

\begin{figure}[tbp]
\centering % \begin{center}/\end{center} takes some additional vertical space
\includegraphics[width=.49\textwidth,clip]{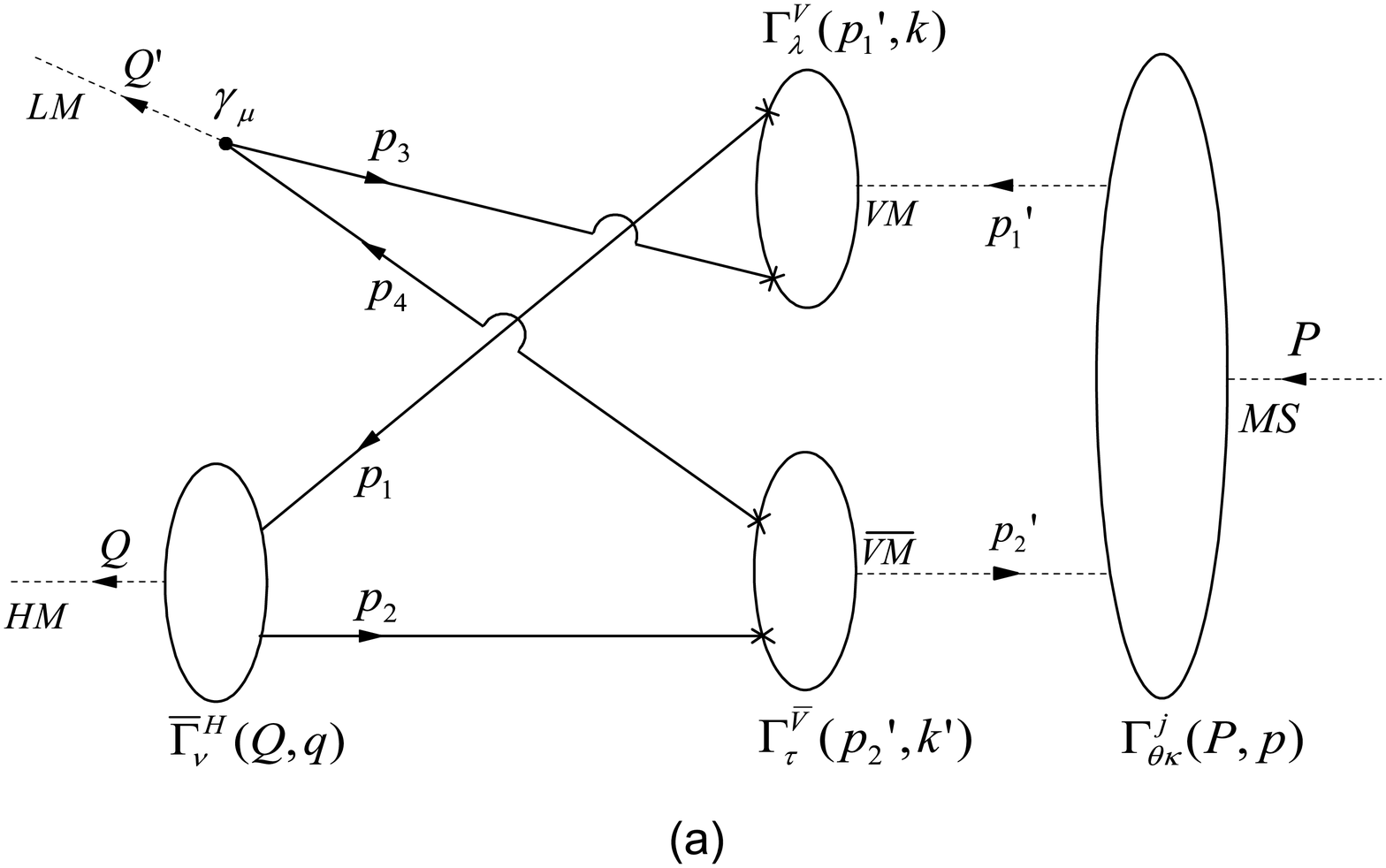}
\hfill
\includegraphics[width=.49\textwidth,origin=c]{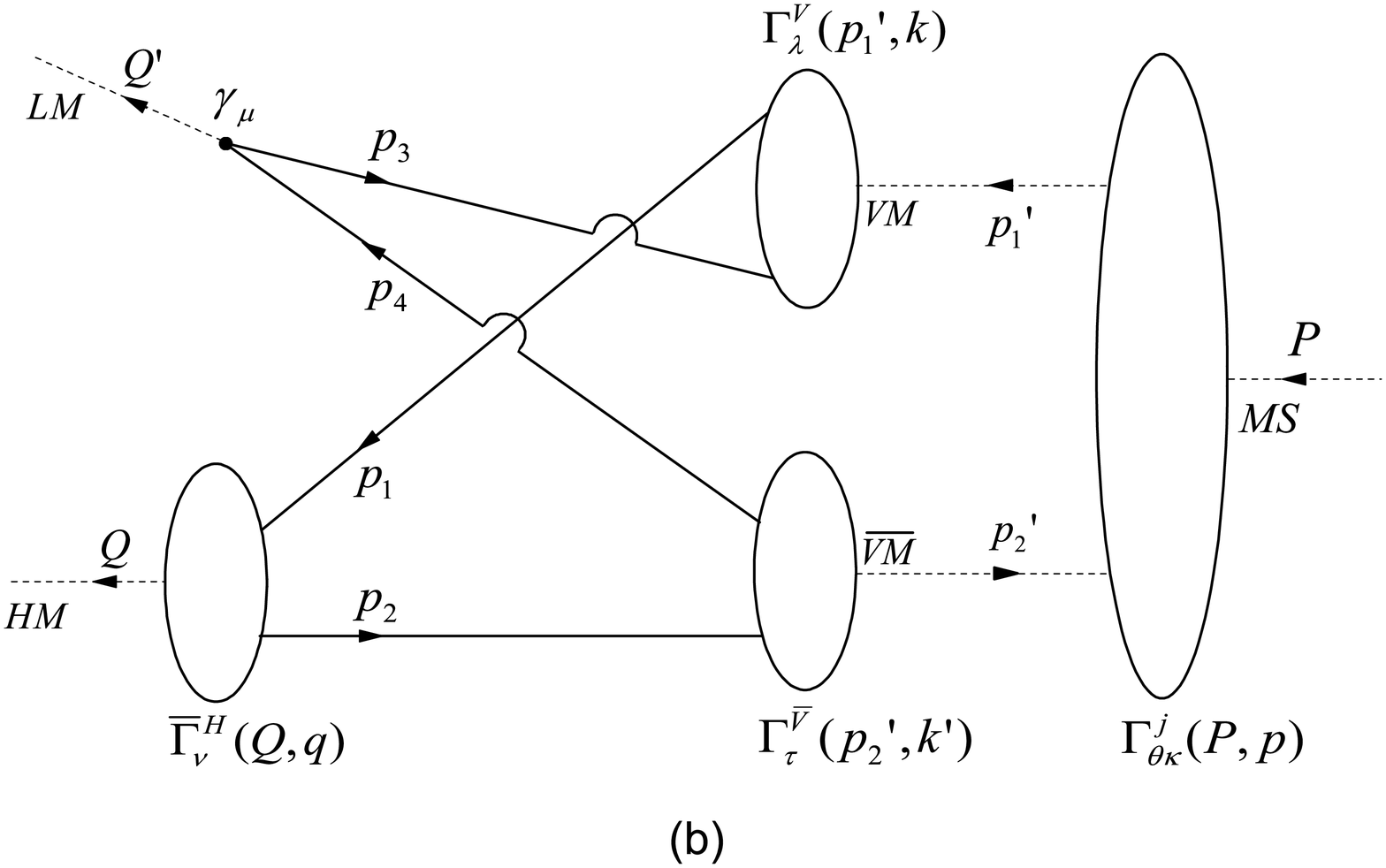}
% "\includegraphics" is very powerful; the graphicx package is already loaded
\caption{\label{Fig2} Matrix element between bound states in the momentum representation. The momenta in the final state satisfy $Q+Q'=P^{\epsilon'}$ and the momentum of the initial state is $P$. The solid lines denote quark propagators. In the rest frame, we have $P=(0,0,0,iM_0)$, $P^{\epsilon'}=(0,0,0,i\epsilon')$ and $\epsilon_M<\epsilon'<\infty$. In diagram (a) the final state energy extends from $\epsilon_{M}$ to $+\infty$ while the initial state energy is specified, and the crosses mean that the momenta of quark propagators depend on the final state energy; in diagram (b) the crosses disappear when $\epsilon'= M_0$.}
\end{figure}

Taking into account the internal structure of heavy mesons ($VM$, $\overline{VM}$ and $HM$) and retaining the lowest order value of $\mathbb{T}$ \cite{mypaper6}, we can obtain the $T$-matrix element with respect to arbitrary value of the final state energy $\epsilon'$ in the momentum representation
\begin{equation}\label{Tmatrele}
\begin{split}
T_{(c';b)a}(\epsilon')=&\frac{ig'_0\varepsilon_\mu^{\varrho'}(Q')\varepsilon_\nu^{\varrho}(Q)}{(2\pi)^{9/2}\sqrt{8E_H(Q)E_L(Q')E(P)}}\int \frac{d^4kd^4p}{(2\pi)^8}\text{Tr}[S_F^\mathcal{D}(p_2)\bar\Gamma^H_\nu(Q,q)S_F^\mathcal{C}(p_1)\\
&\times\Gamma^V_\lambda(p_1',k)S_F^\mathcal{A}(p_3)\gamma_\mu S_F^\mathcal{B}(p_4)\Gamma^{\bar V}_\tau(p_2',k')\chi^j_{\lambda\tau}(P,p)],
\end{split}
\end{equation}
where $p_1, p_3, p_4, p_2$ are the momenta of four quarks; $p_1'$ and $p_2'$ are the momenta of two heavy vector mesons; $q$, $k$ and $k'$ are the relative momenta between quark and antiquark in heavy mesons, respectively; $\varepsilon(p)$ is the polarization vector of vector meson with momentum $p$, $\Gamma^H(K,k)$ represents BS amplitude of heavy meson, $S_F(p)$ is the quark propagator and its superscript is a flavor label, shown as Figure \ref{Fig2}. In our approach, the initial bound state is considered as a four-quark state \cite{mypaper6}, so the generalized BS amplitude of initial bound state should be $\Gamma^V_\lambda(p_1',k)\chi^j_{\lambda\tau}(P,p)\Gamma^{\bar V}_\tau(p_2',k')$, which has been specified. In Figure \ref{Fig2}(a), the energy in $\mathbb{T}$ is equal to the energy of final state, and then the quark momenta in left-hand side of crosses depend on the final state energy and the momenta in right-hand side depend on the initial state energy, i.e., $p_1-p_2-p_3+p_4=Q+Q'=P^{\epsilon'}$ and $p_1'-p_2'=P$. In the rest frame, we have $P=(0,0,0,iM_0)$, $P^{\epsilon'}=(0,0,0,i\epsilon')$ and $\epsilon_M<\epsilon'<\infty$. When $\epsilon'=M_0$, the crosses in Figure \ref{Fig2}(a) disappear and Figure \ref{Fig2}(a) becomes the traditional Feynman diagram, shown as Figure \ref{Fig2}(b). From Figure \ref{Fig2}(b), we have calculated the matrix element $T_{(c';b)a}(M_0)$ and the decay width $\Gamma(M_0)$ with the mass of meson-meson bound state \cite{mypaper6}.

In general, the decay width of resonance is very small compared with its energy level, i.e., $(2\pi)^3\mathbb{I}(M_0)\ll M_0$. This situation is ordinarily interpreted as implying that both $(2\pi)^3|\mathbb{D}(\epsilon)|$ and $(2\pi)^3\mathbb{I}(\epsilon)$ are also very small quantities, as compared to $M_0$. Finally, we can expect that $G_{aa}(\epsilon)$ has a pole on the second Riemann sheet from Eq. (\ref{Gaepsilon})
\begin{equation}
\begin{split}\label{pole}
\epsilon_{\text{pole}}\cong M_0+(2\pi)^3[\mathbb{D}(M_0)-i\mathbb{I}(M_0)]=M-i\Gamma(M_0)/2,
\end{split}
\end{equation}
where $\Delta M=(2\pi)^3\mathbb{D}(M_0)$ is the correction for energy level of resonance and $M=M_0+(2\pi)^3\mathbb{D}(M_0)$ is the physical mass for resonance. The mass $M_0$ of two-body bound state is obtained by solving homogeneous BS equation (\ref{BSE}), which should not be the mass of physical resonance. $\Gamma(M_0)$ with mass $M_0$ also should not be the width of physical resonance, which should depend on its physical mass $M$. Using the mass $M_0$ and BS wave function of bound state, we can calculate the imaginary part $\mathbb{I}(M_0)$ and obtain the correction $\Delta M=(2\pi)^3\mathbb{D}(M_0)$ due to decay channels. Replacing $M_0$ in the momentum of initial bound state by $M$ and setting $\epsilon'=M$, we can calculate the matrix element $T_{(c';b)a}(M)$ from Eq. (\ref{Rmatrix}) and obtain the decay width $\Gamma$ for physical resonance.

Up to now, a theoretical approach to investigate resonance in the framework of relativistic quantum field theory has been established, which is a new way based on BS theory. In this paper, we only explore exotic meson resonance which is considered as an unstable meson-meson molecular state. The extension of our approach to more general resonances is straightforward, while the interaction Lagrangian may be modified.

\section{Example}
As an illustration, we investigate exotic state $\chi_{c0}(3915)$ \cite{Y39401,X39153,X39152}, once named $\emph{X}(3915)$. In experiments two strong decay modes of $\chi_{c0}(3915)$ have been observed: $J/\psi\omega$ and $D^+D^-$. Here, we assume that the isoscalar $\chi_{c0}(3915)$ is a mixed state of two unstable molecular states $D^{*0}\bar{D}^{*0}$ and $D^{*+}D^{*-}$ with spin-parity quantum numbers $0^+$. Firstly, we consider the mixed state of two bound states $D^{*0}\bar{D}^{*0}$ and $D^{*+}D^{*-}$, which can be denoted as $1/\sqrt{2}|D^{*0}\bar{D}^{*0}\rangle+1/\sqrt{2}|D^{*+}D^{*-}\rangle$. In Refs. \cite{mypaper4,mypaper7,mypaper5}, we have obtained the mass $M_0$ and BS wave function $\chi^{0^+}_{\lambda\tau}(P,p)$ for this mixed state of two bound states $D^{*0}\bar{D}^{*0}$ and $D^{*+}D^{*-}$ without an adjustable parameter. In this paper, our attention is focused on the mass correction due to all decay channels and the decay width of physical resonance.

Let us list all open and closed channels. The narrow state $\chi_{c0}(3915)$ was discovered in 2005 \cite{Y39401} and for a long time a series of experiments only observed one strong decay mode of $\chi_{c0}(3915)$: $J/\psi\omega$ denoted as $c'_1$. In 2020 LHCb Collaboration observed another decay channel $D^+D^-$ \cite{X39152} denoted as $c'_2$. Though the neutral channel $D^0\bar D^0$ still has not been observed, this neutral channel should exist for the isospin conservation, which is denoted as $c'_3$. Because the total energy $\epsilon'$ of the final state extends from $\epsilon_M$ to $+\infty$, we obtain one closed channel $D^{*}\bar D^{*}$ derived from the interaction Lagrangian (\ref{Lag}), denoted as $c'_4$. Since bound state lies below the threshold, i.e., $M_0<M_{D^*}+M_{\bar D^*}$, the closed channel $c'_4$ can not occur inside the physical world. Then we can apply Eqs. (\ref{Rmatrix}) and (\ref{Gmatrele}) to evaluate the $T$-matrix element $T_{(c';b)a}(\epsilon')$ for arbitrary decay channel.

In Figure \ref{Fig2}, $VM$ and $\overline{VM}$ become $D^{*}$ and $\bar{D}^{*}$, respectively; $HM$ becomes $J/\psi$ with momentum $Q$ ($Q^2=-M_{J/\psi}^2$) and $LM$ becomes $\omega$ with momentum $Q'$ ($Q'^2=-M_{\omega}^2$); and decay channel $J/\psi\omega$ can be exhibited by these two Feynman diagrams. Here, we consider that the light vector meson $\omega$ is an elementary particle and the heavy vector meson $J/\psi$ is a bound state of $c\bar c$. $\varepsilon_\nu^{\varrho=1,2,3}(Q)$ and $\varepsilon_\mu^{\varrho'=1,2,3}(Q')$ represent the polarization vectors of $J/\psi$ and $\omega$, respectively. Applying Eq. (\ref{Iepsilon'}), we obtain the function
\begin{equation}
\begin{split}\label{Iepsilon'1}
\mathbb{I}_1(\epsilon')=&\frac{1}{2}\int d^3Qd^3Q'(2\pi)^4\delta^{(4)}(Q+Q'-P^{\epsilon'})\sum_{\varrho'=1}^3\sum_{\varrho=1}^3|T_{(c'_1;b)a}(\epsilon')|^2,
\end{split}
\end{equation}
where $T_{(c'_1;b)a}(\epsilon')$ is the bound state matrix element with respect to $\epsilon'$. These heavy mesons $D^{*}$ and $\bar D^{*}$ are considered as quark-antiquark bound states, and $T_{(c'_1;b)a}(\epsilon')$ has been given by Eq. (\ref{Tmatrele}), where flavor labels $\mathcal{C}=\mathcal{D}$ and $\mathcal{A}=\mathcal{B}$ represent $c$-quark and light quark, respectively. The meson-quark coupling constant $g'_0$ becomes $g_\omega$ and $g_\omega^2=2.42/2$ was obtained within QCD sum rules approach \cite{cc2}. BS amplitudes of heavy vector mesons $J/\psi$ and $D^{*}$ have the form $\Gamma_\lambda^V(K,k)=(\gamma_\lambda+K_\lambda\gamma\cdot K/M_V^2)\text{exp}(-k^2/\omega_V^2)$, where $\omega_{J/\psi}$=0.826GeV \cite{mypaper6,Jpsi1} and $\omega_{D^{*}}$=1.50GeV \cite{BSE:Roberts5}. These momenta in Figure \ref{Fig2}(a) become $p_1=(Q+Q')/2+p+k$, $p_2=(Q'-Q)/2+p+k$, $p_3=k$, $p_4=Q'+k$, $p_1'=p+P/2$, $p_2'=p-P/2$, $Q+Q'=P^{\epsilon'}=(0,0,0,i\epsilon')$ and $P=(0,0,0,iM_0)$. Using Eq. (\ref{Tmatrele}), we can calculate the $T$-matrix element $T_{(c'_1;b)a}(\epsilon')$ with respect to arbitrary energy $\epsilon'$ for channel $c'_1$. From Eq. (\ref{Iepsilon'1}), we obtain the function $\mathbb{I}_1(\epsilon')$ for channel $c'_1$ and dispersion relation (\ref{disrel}) becomes
\begin{equation}
\begin{split}
\mathbb{D}_1(M_0)=-\frac{\mathcal{P}}{\pi}\int_{\epsilon_{c'_1,M}}^\infty \frac{\mathbb{I}_1(\epsilon')}{\epsilon'-M_0}d\epsilon',
\end{split}
\end{equation}
where $\epsilon_{c'_1,M}=M_{J/\psi}+M_\omega$. The $T$-matrix element $T_{(c'_1;b)a}(\epsilon')$ and the function $\mathbb{I}_1(\epsilon')$ for channel $c'_1$ are calculated over the real interval $\epsilon_{c'_1,M}<\epsilon'<\infty$, and we obtain the mass correction $\Delta M_1=(2\pi)^3\mathbb{D}_1(M_0)$ due to channel $c'_1$.

\begin{figure}[tbp] \centering
\includegraphics[trim = 0mm 40mm 0mm 45mm,scale=1,width=10.5cm]{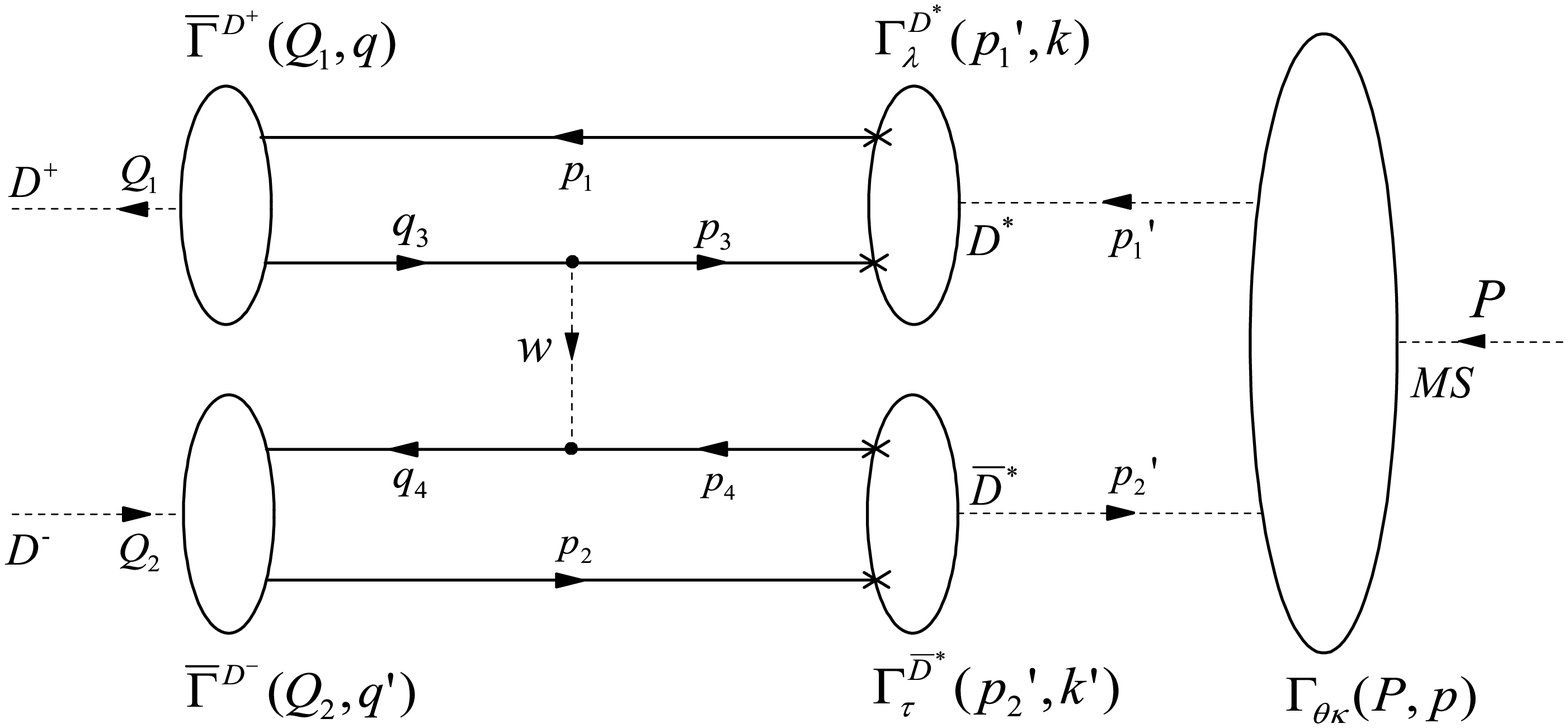}%trim=l b r t This option will crop the imported image by l from the left, b from the bottom, r from the right, and t from the top. Where l, b, r and t are lengths.
\caption{\label{Fig3} Matrix element for decay channel $D^+D^-$. The momenta in the final state satisfy $Q_1+Q_2=P^{\epsilon'}$, the momentum of the initial state is $P$, and we have $P=(0,0,0,iM_0)$, $P^{\epsilon'}=(0,0,0,i\epsilon')$ and $\epsilon_{c'_2,M}<\epsilon'<\infty$. $w$ represents the momentum of the exchanged light meson. The crosses mean that the momenta of quark propagators and the momentum $w$ of the exchanged light meson depend on the final state energy.}
\end{figure}

For decay channel $D^{+}D^{-}$, we consider that these heavy pseudoscalar mesons $D^{+}$ and $D^{-}$ are quark-antiquark bound states. $Q_1$ and $Q_2$ represent the momenta of final particles, $Q_1^2=-M_{D^{+}}^2$ and $Q_2^2=-M_{D^{-}}^2$. From Eq. (\ref{Iepsilon'}), we obtain the function $\mathbb{I}_2(\epsilon')$
\begin{equation}
\begin{split}\label{Iepsilon'2}
\mathbb{I}_2(\epsilon')=&\frac{1}{2}\int d^3Q_1d^3Q_2(2\pi)^4\delta^{(4)}(Q_1+Q_2-P^{\epsilon'})|T_{(c'_2;b)a}(\epsilon')|^2,
\end{split}
\end{equation}
where $T_{(c'_2;b)a}(\epsilon')$ represents the bound state matrix element with $\epsilon'$. Considering the lowest order term of $\mathbb{T}$, we obtain $T_{(c'_2;b)a}(\epsilon')$ represented graphically by Figure \ref{Fig3}, where $p_1-p_2-p_3+p_4=p_1-p_2-q_3+q_4=Q_1+Q_2=P^{\epsilon'}$, $p_1'-p_2'=P$, and the crosses mean that the momenta of quark propagators and the momentum $w$ of the exchanged light meson depend on $Q_1$ and $Q_2$. The $T$-matrix element with respect to $\epsilon'$ for channel $c'_2$ becomes
\begin{equation}\label{Tmatrele2}
\begin{split}
T_{(c'_2;b)a}(\epsilon')=&\frac{-ig^2}{(2\pi)^{9/2}\sqrt{8E_{D^+}(Q_1)E_{D^-}(Q_2)E(P)}}\int\frac{d^4kd^4k'd^4p}{(2\pi)^{12}}\text{Tr}[S^d_F(q_3)\\
&\times \bar\Gamma^{D^+}(Q_1,q)S^c_F(p_1)\Gamma^{D^{*}}_\lambda(p_1',k)S^l_F(p_3)\mathcal{O}^{dl}(p_3,q_3)]\chi^{0^+}_{\lambda\tau}(P,p)\Delta_F(w)\\
&\times \text{Tr}[S^l_F(p_4)\Gamma^{\bar D^{*}}_\tau(p_2',k')S^c_F(p_2)\bar\Gamma^{D^-}(Q_2,q')S^d_F(q_4)\mathcal{O}^{dl}(q_4,p_4)],
\end{split}
\end{equation}
where $q$, $q'$, $k$ and $k'$ are the relative momenta between quark and antiquark in heavy mesons, respectively; the meson-quark coupling constants $g$ were obtained within QCD sum rules approach \cite{mypaper3,mypaper4,cc2}, $\mathcal{O}^{dl}(p,q)$ represents the meson-quark vertex, $\Delta_F(w)$ is the light meson propagator, the superscript of quark propagator $S_F(p)$ is flavor label, and $l=u,d$ represents the $u,d$-antiquark in heavy vector meson $D^{*0}$ or $D^{*+}$, respectively. BS amplitude of heavy pseudoscalar meson $D^{+}$ has the form $\Gamma^{D^+}(K,k)=i\gamma_5\text{exp}(-k^2/\omega_{D}^2)$, where $\omega_{D}$=1.50GeV \cite{BSE:Roberts5}. The meson-quark vertex $\mathcal{O}^{dl}(p,q)$ is unit matrix for one-$\sigma$ exchange; and it becomes $\gamma_\mu$ for one light vector meson exchange. From Eqs. (\ref{Iepsilon'2}), (\ref{disrel}) and (\ref{pole}), we obtain the mass correction $\Delta M_2$ due to channel $c'_2$
\begin{equation}
\begin{split}
\Delta M_2=(2\pi)^3\mathbb{D}_2(M_0)=-\frac{\mathcal{P}}{\pi}\int_{\epsilon_{c'_2,M}}^\infty \frac{(2\pi)^3\mathbb{I}_2(\epsilon')}{\epsilon'-M_0}d\epsilon',
\end{split}
\end{equation}
where $\epsilon_{c'_2,M}=M_{D^+}+M_{D^-}$. The $T$-matrix element $T_{(c'_2;b)a}(\epsilon')$ and the function $\mathbb{I}_2(\epsilon')$ for channel $c'_2$ are calculated over the real interval $\epsilon_{c'_2,M}<\epsilon'<\infty$. Following the same procedure as for channels $c'_1$ and $c'_2$, we can calculate the mass corrections $\Delta M_3$ and $\Delta M_4$ due to the channel $c'_3$ and closed channel $c'_4$, respectively.

Considering the isospin conservation, we employ the constituent quark masses $m_u=m_d=0.33$GeV, the heavy quark mass $m_{c}=1.55$GeV \cite{ts:Ebert} and the meson masses $M_\sigma=0.45$GeV, $M_{\omega}=0.782$GeV, $M_{\rho}=0.775$GeV, $M_\phi=1.019$GeV, $M_{D^{*0}}=M_{D^{*+}}=2.007$GeV, $M_{D^{0}}=M_{D^{+}}=1.865$GeV, $M_{J/\psi}=3.097$GeV \cite{PDG2022}. By doing the numerical calculation, we obtain the mass corrections $\Delta M_i(i=1,2,3,4)$ due to three open decay channels $J/\psi\omega$, $D^+D^-$, $D^0\bar D^0$ and one closed channel $D^{*}\bar D^{*}$, respectively. Subsequently, the mass $M$ for physical resonance $\chi_{c0}(3915)$ can be applied to calculate its decay width. Replacing $M_0$ by $M$ in Eq. (\ref{Tmatrele}) and setting $\epsilon'=M$, we calculate the matrix element $T_{(c'_1;b)a}(M)$ and obtain that the width for physical decay model $\chi_{c0}(3915)\rightarrow J/\psi\omega$ is $\Gamma_1=2(2\pi)^3\mathbb{I}_1(M)$. Replacing $M_0$ by $M$ in Eq. (\ref{Tmatrele2}) and setting $\epsilon'=M$, we calculate the matrix element $T_{(c'_2;b)a}(M)$ and obtain that the width for physical decay model $\chi_{c0}(3915)\rightarrow D^+D^-$ is $\Gamma_2=2(2\pi)^3\mathbb{I}_2(M)$. For the isospin conservation, it is easy to obtain the width $\Gamma_3$ for physical decay model $\chi_{c0}(3915)\rightarrow D^0\bar D^0$. Our numerical results are presented in Table \ref{table1}, and the mass $M$ and full width $\Gamma$ are in good agreement with experimental data. Furthermore, the calculated $D^+D^-$ width $\Gamma_2$ is very small compared with the calculated $J/\psi\omega$ width $\Gamma_1$, and then we can explain why the decay model $\chi_{c0}(3915)\rightarrow D^+D^-$ had not been observed in experiments for a long time. Therefore, this work provides a further verification for the molecular hypothesis of $\chi_{c0}(3915)$ and predicts the exact values of these strong decay widths $\Gamma_1(\chi_{c0}(3915)\rightarrow J/\psi\omega)$, $\Gamma_2(\chi_{c0}(3915)\rightarrow D^+D^-)$ and $\Gamma_3(\chi_{c0}(3915)\rightarrow D^0\bar D^0)$.
\newcommand{\tabincell}[2]{\begin{tabular}{@{}#1@{}}#2\end{tabular}}
\begin{table}[tbp]
\vspace*{-6pt}
\tabcolsep=1.5mm
%\small
\begin{center}
\begin{tabular}{|c|c|c|c|c|c|c|c|c|c|c|}\hline
Quantity  & $M_0$ & $\Delta M_1$  & $\Delta M_2$ & $\Delta M_3$ & $\Delta M_4$ & $M$ & $\Gamma_1$ & $\Gamma_2$ & $\Gamma_3$ &$\Gamma$
\\ \hline
this work     & 3953.7 & $-$24.0 & $-$1.4 & $-$1.4 & $-$4.6 &  3922.3  & 22.3 & 1.5 &1.5& 25.3
\\
\hline
PDG\cite{PDG2022} & & & & & & 3921.7$\pm$1.8 & & & & 18.8$\pm$3.5
\\
\hline
\end{tabular}
\caption{\label{table1} Mass $M$ and width $\Gamma$ for physical resonance $\chi_{c0}(3915)$. $M_0$ is the mass of mixed state of two bound states $D^{*0}\bar{D}^{*0}$ and $D^{*+}D^{*-}$, $\Delta M_i$ is the calculated correction due to $i$th decay channel. (Dimensioned quantities in MeV.)}
\end{center}
\end{table}

In the actual calculation, we require the meson-quark coupling constants $g$ and the parameters $\omega_H$ in BS amplitudes of heavy mesons to calculate the mass and decay width of physical resonance. The meson-quark coupling constants can be determined by QCD sum rules approach \cite{cc2}, and these parameters in BS amplitudes of heavy mesons are fixed by providing fits to observables \cite{BSE:Roberts4,BSE:Roberts5,Jpsi1}. Our approach also involves the constituent quark masses $m_u$, $m_d$, and the heavy quark mass $m_c$. According to the spontaneous breaking of chiral symmetry, the light quarks ($u,d,s$) obtain their constituent masses because the vacuum condensate is not equal to zero, and the heavy quark mass $m_c$ is irrelevant to vacuum condensate. Normally, the value slightly greater than a third of nucleon mass is employed as the constituent mass of light quark. The value of heavy quark mass $m_c$ can be determined by the experimental mass of charmonium system $J/\psi$. Therefore, there is not an adjustable parameter in our approach. Of course, the values of these parameters, including $g$, $\omega_H$, $m_u$, $m_d$ and $m_c$, are values in respective ranges. Simultaneously varying these parameters in respective ranges, we find that the uncertainties of numerical results are at most 5\%. Despite the large uncertainty of meson mass $M_\sigma$, it has been found that the uncertainties of numerical results from meson mass $M_\sigma$ are also very small in our previous works \cite{mypaper4,mypaper5,mypaper6,mypaper7} and Refs. \cite{Msigma1,Msigma2}. Thus in our approach the calculated mass and decay width are uniquely determined. In this paper we emphatically illuminate the physical meaning of new resonance theory in quantum field theory, and the details in computational process will be shown in our future article.

\section{Conclusion}
We recognize that resonance can not be completely treated as a stationary bound state and provide a reasonable and feasible scheme to describe unstable system in the framework of relativistic quantum field theory. Based on BS wave function, we provide a description of the prepared state and investigate the time evolution of two-body bound state as determined by the total Hamiltonian. According to dispersion relation, the total matrix elements for all decay channels should be calculated with respect to arbitrary value of the final state energy, and these matrix elements are expressed in terms of the Heisenberg picture. Mandelstam's approach is generalized to calculate the matrix element between bound states with arbitrary value of the final state energy, which is exhibited in extended Feynman diagram. Finally, the mass and decay width for physical resonance are obtained. In this paper, we illustrate this new resonance theory in quantum field theory by reference to the example of exotic meson which is considered as an unstable meson-meson molecular state, and obviously our work can be extended to more general resonances and creates a new paradigm for investigating hadron resonances.

\acknowledgments
This work was supported by the National Natural Science Foundation of China under Grants No. 11705104, No. 11801323 and No. 52174145; Shandong Provincial Natural Science Foundation, China under Grants No. ZR2016AQ19 and No. ZR2016AM31; and SDUST Research Fund under Grant No. 2018TDJH101.


%\bibliography{ref}

\begin{thebibliography}{99}

\bibitem{ms:Swanson}
E.S. Swanson, \emph{Short range structure in the $X(3872)$}, \emph{Phys. Lett. B} {\bf 588} (2004) 189.

\bibitem{ms:Torn}
N.A. T{\"o}rnqvist, \emph{Isospin breaking of the narrow charmonium state of Belle at 3872 MeV as a deuson}, \emph{Phys. Lett. B} {\bf 590} (2004) 209.

\bibitem{liu}
X. Liu, S.L. Zhu, \emph{$Y(4143)$ is probably a molecular partner of $Y(3930)$}, \emph{Phys. Rev. D} {\bf 80} (2009) 017502.


\bibitem{ds:Maian1}
L. Maiani, F. Piccinini, A.D. Polosa, V. Riquer, \emph{Diquark-antidiquark states with hidden or open charm and the nature of $X(3872)$}, \emph{Phys. Rev. D} {\bf 71} (2005) 014028.

\bibitem{ds:Maian2}
L. Maiani, A.D. Polosa, V. Riquer, \emph{Indications of a Four-Quark Structure for the $X(3872)$ and $X(3876)$ Particles from Recent Belle and BABAR Data}, \emph{Phys. Rev. Lett.} {\bf 99} (2007) 182003.

\bibitem{ts:Ebert}
D. Ebert, R.N. Faustov, V.O. Galkin, \emph{Masses of heavy tetraquarks in the relativistic quark model}, \emph{Phys. Lett. B} {\bf 634} (2006) 214.

\bibitem{ms:Branz}
T. Branz, T. Gutsche, V.E. Lyubovitskij, \emph{Hadronic molecule structure of the $Y(3940)$ and $Y(4140)$}, \emph{Phys. Rev. D} {\bf 80} (2009) 054019.

\bibitem{mypaper4}
X. Chen, X. L{\"u}, \emph{Mass of $Y(3940)$ in Bethe-Salpeter equation for quarks}, \emph{Eur. Phys. J. C} {\bf 75} (2015) 98.

\bibitem{mypaper6}
X. Chen, X. L\"u, \emph{Decay width of hadronic molecule structure for quarks}, \emph{Phys. Rev. D} {\bf 97} (2018) 114005.

\bibitem{mypaper7}
X. Chen, X. L\"u, R. Shi, X. Guo, Q. Wang, \emph{Radiative decay of hadronic molecule state for quarks}, \emph{Phys. Rev. D} {\bf 101} (2020) 014009.

\bibitem{Msigma2}
M.-J. Zhao, Z.-Y. Wang, C. Wang, X.-H. Guo, \emph{Investigation of the possible $D{\bar{D}}^{*}/B{\bar{B}}^{*}$
	and $D{D}^{*}/\bar{B}{\bar{B}}^{*}$ bound states}, \emph{Phys. Rev. D} {\bf 105}, (2022) 096016.

\bibitem{BSE:Roberts3}
P. Maris, C.D. Roberts, P.C. Tandy, \emph{Pion mass and decay constant}, \emph{Phys. Lett. B} {\bf 420} (1998) 267.

\bibitem{BSE:Roberts4}
M.A. Ivanov, Y.L. Kalinovsky, C.D. Roberts, \emph{Survey of heavy-meson observables}, \emph{Phys. Rev. D} {\bf 60} (1999) 034018.

\bibitem{BSE:Roberts5}
M.A. Ivanov, J.G. K\"orner, S.G. Kovalenko, C.D. Roberts, \emph{$B$-meson to light-meson transition form factors}, \emph{Phys. Rev. D} {\bf 76} (2007) 034018.

\bibitem{couchan1}
C. Chin, R. Grimm, P. Julienne, E. Tiesinga, \emph{Feshbach resonances in ultracold gases}, \emph{Rev. Mod. Phys.} {\bf 82}, (2010) 1225.

\bibitem{couchan2}
M.R. Pennington, D.J. Wilson, \emph{Decay channels and charmonium mass shifts}, \emph{Phys. Rev. D} {\bf 76}, (2007) 077502.

\bibitem{couchan3}
M.-X. Duan, S.-Q. Luo, X. Liu, T. Matsuki, \emph{Possibility of charmoniumlike state $X(3915)$ as ${\ensuremath{\chi}}_{c0}(2P)$ state}, \emph{Phys. Rev. D} {\bf 101}, (2020) 054029.

\bibitem{QTFII}
S. Weinberg, \emph{The Quantum Theory of Fields}, vol.~\bibinfo{volume}{\textrm{II}}, Cambridge University Press, (1996).

\bibitem{mypaper5}
X. Chen, X. L{\"u}, R. Shi, X. Guo, \emph{Calculation of mass of $Y(4140)$ by introducing mixed molecule state in quark model}, \emph{Nucl. Phys. B} {\bf 909} (2016) 243.

\bibitem{GreenFun}
M.L. Goldberger, K.M. Watson, \emph{Collision Theory}, Wiley, New York, (1964).

\bibitem{ParticlesFields}
D. Luri\'{e}, \emph{Particles and Fields}, Interscience Publishers, New York, (1968).

\bibitem{mypaper3}
X. Chen, R. Liu, R. Shi, X. L\"u, \emph{Bethe-Salpeter wave functions for the bound states composed of two vector fields of arbitrary spin and their application}, \emph{Phys. Rev. D} {\bf 87} (2013) 065013.

\bibitem{Y39401}
S.-K. Choi, et al., (Belle Collaboration), \emph{Observation of a near-threshold $\ensuremath{\omega}J/\ensuremath{\psi}$ mass enhancement in exclusive $B\ensuremath{\rightarrow}K\ensuremath{\omega}J/\ensuremath{\psi}$ decays}, \emph{Phys. Rev. Lett.} {\bf 94} (2005) 182002.

\bibitem{X39153}
A. Vinokurova, et al., (Belle Collaboration), \emph{Search for B decays to final states with the $\eta_c$ meson}, \emph{JHEP} {\bf 06} (2015) 132.

\bibitem{X39152}
R. Aaij, et al., (LHCb Collaboration), \emph{Amplitude analysis of the ${B}^{+}\ensuremath{\rightarrow}{D}^{+}{D}^{\ensuremath{-}}{K}^{+}$ decay}, \emph{Phys. Rev. D} {\bf 102} (2020) 112003.

\bibitem{cc2}
L. Reinders, H. Rubinstein, S. Yazaki, \emph{Hadron properties from QCD sum rules}, \emph{Phys. Rep.} {\bf 127}, (1985) 1.

\bibitem{Jpsi1}
T. Kawanai, S. Sasaki, \emph{Interquark Potential with Finite Quark Mass from Lattice QCD}, \emph{Phys. Rev. Lett.} {\bf 107}, (2011) 091601.

\bibitem{PDG2022}
R.L. Workman, et al., (Particle Data Group), \emph{Review of particle physics}, \emph{Prog. Theor. Exp. Phys.} {\bf 2022} (2022) 083C01.

\bibitem{Msigma1}
G.-J. Ding, \emph{Are $Y(4260)$ and ${Z}_{2}^{+}(4250)$ ${D}_{1}D$ or ${D}_{0}{D}^{*}$
	hadronic molecules?}, \emph{Phys. Rev. D} {\bf 79}, (2009) 014001.

\end{thebibliography}

\end{document}